\documentclass[12pt]{article}
\usepackage{graphicx}

\newcommand \bra[1]{\left< {#1} \,\right\vert}
\newcommand \ket[1]{\left\vert\, {#1} \, \right>}

\newcommand{\bea}{\begin{eqnarray}}
\newcommand{\eea}{\end{eqnarray}}
\newcommand{\simgt}{\hbox{ \raise3pt\hbox to 0pt{$>$}\raise-3pt\hbox{$\sim$} }}
\newcommand{\simlt}{\hbox{ \raise3pt\hbox to 0pt{$<$}\raise-3pt\hbox{$\sim$} }}

\begin{document}
\begin{titlepage}
\title{A Connection Between The Perturbative QCD Potential\\
and Phenomenological Potentials \vspace{2cm}}
\author{Y.~Sumino
\\ \\ Department of Physics, Tohoku University\\
Sendai, 980-8578 Japan
}
\date{}
\maketitle
\thispagestyle{empty}
\vspace{-4truein}
\begin{flushright}
{\bf TU--618}\\
{\bf April 2001}
\end{flushright}
\vspace{4.5truein}
\begin{abstract}
\noindent
{\small
When the cancellation of the leading renormalon 
contributions is incorporated, the total energy of a $b\bar{b}$ system,
$E_{{\rm tot},b\bar{b}}(r) \equiv 2 \, m_{{\rm pole},b} + V_{\rm QCD}(r)$,
agrees well with the potentials used in phenomenological
models for heavy quarkonia in the range 
$0.5~{\rm GeV}^{-1} \simlt r \simlt 3~{\rm GeV}^{-1}$.
We provide a connection between the conventional 
potential-model approaches to the quarkonium spectroscopy
and the recent computation based on perturbative QCD.
}
\end{abstract}
\vfil

\end{titlepage}
  
\section{Introduction}
\label{intro}

For over 20 years, most successful theoretical approaches for describing
the charmonium and bottomonium systems (including the excited states) 
have been those based on
various phenomenological potential models.
These phenomenological-model approaches have elucidated
nature of the heavy quarkonium systems, such as their leptonic widths
and transitions among different levels, besides reproducing
the energy levels.
The phenomenological potentials determined and used in these studies
have more or less similar slopes in the range 
$0.5~{\rm GeV}^{-1} \simlt r \simlt$ $5~{\rm GeV}^{-1}$, which
may be represented by a logarithmic potential
$\propto \log r + {\rm const}$.
See e.g.\ Ref.\cite{eq} for a recent analysis based on the
potential models.
An apparent deficit of these approaches is, however, a
difficulty in relating phenomenological parameters to the fundamental
parameters of QCD.

The reason why people have been using phenomenological models is
because the theory of non-relativistic boundstates based on
perturbative QCD
failed to reproduce the charmonium and bottomonium spectra.
This is in contrast to the corresponding theory based on perturbative QED,
which has been successful in describing the spectra of the QED boundstates.
The main problem has been poor convergence of the perturbative
expansions when the energy levels of the heavy quarkonia 
are computed in series expansions in the strong coupling constant.
Since the coupling constant is quite large at relevant scales, approximating
order one, it has been considered as an indication of
large non-perturbative effects inherent in these quarkonium systems.
In fact the difference between a typical phenomenological potential
and the Coulomb potential tends to be a linearly rising potential at
distances $r \simgt 1~{\rm GeV}^{-1}$,
suggesting confinement of quarks.
Within perturbative QCD, the origin of the poor convergence has been
understood in terms of the renormalon contributions \cite{al}.

More recently, theoretical frameworks based on QCD
have been developed for describing these quarkonium systems 
systematically.
Within effective theories based on appropriate expansions
in small parameters, various potentials are defined 
such that the leading-order potential plays a role
close to that of the potentials introduced in the above phenomenological
approaches.
The order countings of terms in organizing the expansions 
depend crucially on the relative sizes of the dynamically generated scales
(soft scale $\sim m v$, ultrasoft scale $\sim m v^2$, where
$m$ and $v$ are the quark mass and velocity, respectively)
and the hadronization scale $\sim \Lambda_{\rm QCD}$.
This aspect contrasts with the fact that the expansion parameter
in the non-relativistic boundstate theory
based on perturbative QCD is simply $1/c$ (inverse of the speed of light).
In the formalism developed in \cite{bw,ef,gromes,bmp,bbp}, or in
the potential-Non-relativistic QCD (pNRQCD) formalism
\cite{pinedasoto,bpsv} formulated more systematically,
potentials are defined in a way 
suited for practical computations by lattice simulations 
(or by using models).
On the basis of these formalisms,
lattice calculations have shown
from first principles that the leading-order potential has a shape
consistent with the phenomenological potentials in the relevant range,
although the accuracy of the computations needs further improvements
\cite{bsw,ukcollab}.

Very recently, a new computation of the charmonium and bottomonium spectra
has been reported
in the framework of non-relativistic boundstate theory
based on perturbative QCD \cite{bsv1}.
It incorporated recent significant developments in the field:
(1) the full computations of the quarkonium energy levels up to
order $1/c^2$ \cite{yn3,py,my,bcbv};
(2) the cancellation of the leading renormalons
contained in the quark pole mass and the static QCD potential
\cite{renormalon1,renormalon2}.
As a result, convergence property of the
series expansions of the energy levels improved drastically, which enabled
stable perturbative predictions for the levels up to some of the
$n=3$ bottomonium states and the $n=1$ charmonium states
($n$ is the principal quantum number).
Furthermore, the computed spectrum, when averaged over spins, 
reproduced the gross structure
of the observed energy levels of the bottomonium states, within
moderate theoretical uncertainties estimated from the next-to-leading
renormalon contributions. 
It indicates that non-perturbative contributions to the bottomonium
spectrum, in the scheme free from the leading renormalons,
would absorb the next-to-leading renormalon uncertainties of the
perturbative predictions and may be of the size comparable to them.

It is then natural to ask whether there is a connection between
the above phenomenological potential-model approaches (supplemented
by the more systematic frameworks and lattice calculations) 
and the recent computation based on perturbative QCD.
Once this connection is established, we may merge these approaches
and further develop understandings of the charmonium and bottomonium
systems.
For instance, in the perturbative computation, the level splittings
between the $S$-wave and $P$-wave states as well as
the fine splittings among the $nP_j$ states turn out to
be smaller than the corresponding experimental values.
Although the discrepancy is still smaller than the estimated theoretical
uncertainties of the perturbative predictions, it should certainly be
clarified whether they are explained by higher-order perturbative
corrections, or, we need specific non-perturbative effects for
describing them.
On the other hand, the potential approaches have been
successful also in explaining the $S$-$P$ splittings and the fine splittings.
Hence, we expect that
a connection between these theoretical approaches would help 
to clarify origins of the differences of the present
perturbative predictions and the experimental data.

In this paper we focus on the perturbative static QCD potential
up to ${\cal O}(\alpha_S^3)$, since
it dictates the major structures of the quarkonium spectra 
in the perturbative computation up to ${\cal O}(1/c^2)$ \cite{bsv1}.
Taking into account the above key ingredient (2), 
we subtract the leading renormalon contribution from the QCD potential.
Then we compare it with the phenomenologically determined
potentials.
Our comparison also elucidates to which extent 
the perturbative computation of the QCD potential 
[up to ${\cal O}(\alpha_S^3)$, and after subtracting
the leading renormalon] reproduces the results of the
non-perturbative computations.
(We will regard typical phenomenological potentials as representatives of the
lattice results, taking into account consistency of the potentials
determined in both approaches.)

In Sec.~2 we review the theoretical uncertainties from the renormalon
contributions within the context of the large-$\beta_0$ approximation.
In Sec.~3 we analyze the total energy of a quark-antiquark system up
to ${\cal O}(\alpha_S^3)$.
Also the interquark force is analyzed in Sec.~4.
We draw conclusions in Sec.~5.

\section{Renormalons in the Large-$\beta_0$ Approximation}
\label{sec:largebeta0}

The static QCD potential, defined from an expectation value of
the Wilson loop, represents the potential energy of a
static quark-antiquark pair:
\bea
V_{\rm QCD}(r) 
&=&
- \lim_{T \to \infty} \frac{1}{i\,  T} \,\, \log
\frac{ \bra{0} {\rm Tr} \, {\rm P} \, 
\exp \Bigl[ i g_S \oint_\Gamma dx^\mu A_\mu (x) \Bigr]
\ket{0} }
{ \bra{0} {\rm Tr} \, {\bf 1} \ket{0} }
\\
&=& 
- C_F \frac{\alpha_V(1/r)}{r} ,
\eea
where $\Gamma$ is a rectangular loop of spatial extent $r$ and time
extent $T$.
The second line defines the $V$-scheme coupling constant, $\alpha_V(1/r)$,
where $C_F=4/3$.
In perturbative QCD, 
the $V$-scheme coupling constant is calculable in a series expansion
in the coupling constant as\footnote{
From ${\cal O}(\alpha_S^4)$ and beyond, 
the series includes infrared divergences;
the divergences can be circumvented by a resummation of diagrams,
which brings in $\log \alpha_S$ in the series expansion, or,
$\log (\mu_{\rm eff} \, r)$ term when the theory is matched to the 
pNRQCD effective theory \cite{adm,bpsv}.
}
\bea
\alpha_V(1/r) = \alpha_S(\mu) \sum_{n=0}^\infty 
P_n(\log (\mu r)) 
\left( \frac{\alpha_S(\mu)}{4\pi} \right)^n .
\eea
Throughout this paper, $\alpha_S(\mu)$ denotes the strong coupling constant
in the $\overline{\rm MS}$ scheme with $n_l$ active flavors;
$\mu$ is the renormalization scale;
$P_n(L)$ denotes an $n$-th-degree polynomial of $L$.
Although the exact QCD potential $V_{\rm QCD}(r)$ is independent of
the scale $\mu$, at each order of the perturbative expansion
$\mu$-dependences remain.
We keep $\mu$ as a free parameter in this section.
From an analysis of higher-order terms, it has been known \cite{al} that 
the perturbative expansion of $V_{\rm QCD}(r)$ has an uncertainty of order
$\Lambda_{\rm QCD}$,
which is referred to as the renormalon problem.
We first review this property and estimate uncertainties of the
perturbative prediction for the QCD potential.
(See e.g.\ \cite{pro,sumino} for introductory reviews.)

The ``large-$\beta_0$ approximation'' \cite{bb} is an empirically successful
method for analyzing large-order behaviors of physical quantities
in perturbative QCD and renormalon ambiguities inherent in them.
Let us denote by $V_{\beta_0}(r)$ the QCD potential
within this approximation and by
$V^{(n)}_{\beta_0}(r)$ its ${\cal O}(\alpha_s^{n+1})$ term:
\bea
V_{\beta_0}(r) 
= \sum_{n=0}^\infty V^{(n)}_{\beta_0}(r) .
\eea
From the Taylor expansion of the Borel transform of $V_{\beta_0}(r)$,
we can easily compute $V^{(n)}_{\beta_0}(r)$ one by one from the lowest order.
Also the asymptotic form for $n \gg 1$ is determined as
\bea
V^{(n)}_{\beta_0}(r) \sim 
- C_F \, 4 \pi \alpha_S(\mu) \times\frac{\mu \, e^{5/6}}{2\pi^2} \times
\biggl\{ \frac{\beta_0 \alpha_S(\mu)}{2\pi} \biggr\}^n \,
\, n! ,
\label{asympt}
\eea
where $\beta_0=11-2n_l/3$ is the coefficient of the
QCD one-loop beta function.
The above asymptotic behavior is independent of $r$.
It means that, although each term of 
the potential is
a function of $r$, its dominant part for $n \gg 1$ is
only a constant potential which mimics the role of the quark mass
in the determination of the total energy of a quark-antiquark system.
As we raise $n$, first $|V^{(n)}_{\beta_0}(r)|$ decreases due to powers of 
the small $\alpha_S$;
for very large $n$ it increases due to the factorial $n!$.
Around $n_0 = 2\pi/(\beta_0 \alpha_S(\mu))$,
$|V^{(n)}_{\beta_0}(r)|$ becomes smallest.
The size of the term scarcely changes within the range 
$n \in ( n_0 - \sqrt{n_0}, n_0 + \sqrt{n_0} )$.
We may consider the uncertainty of this asymptotic series
as the sum of the terms within this range, since one may 
equally well truncate
the series at order $n_0 - \sqrt{n_0}$ or at order $n_0 + \sqrt{n_0}$
in estimating the ``true value'' of the potential:
\bea
\delta V_{\beta_0}(r) 
\sim 
\left| \sum_{ n = n_0 - \sqrt{n_0} }^{n_0 + \sqrt{n_0}} \, 
V^{(n)}_{\beta_0}(r) \right|
\sim \Lambda \equiv \mu \, \exp \biggl[
- \, \frac{2\pi}{\beta_0 \alpha_S(\mu)} \biggr] .
\eea
The $\mu$-dependence vanishes in this sum, and this leads to the
claimed uncertainty.
\begin{figure}[t]
  \hspace*{\fill}
  \begin{minipage}{5.0cm}\centering
    \hspace*{-2.3cm}
    \includegraphics[width=8cm]{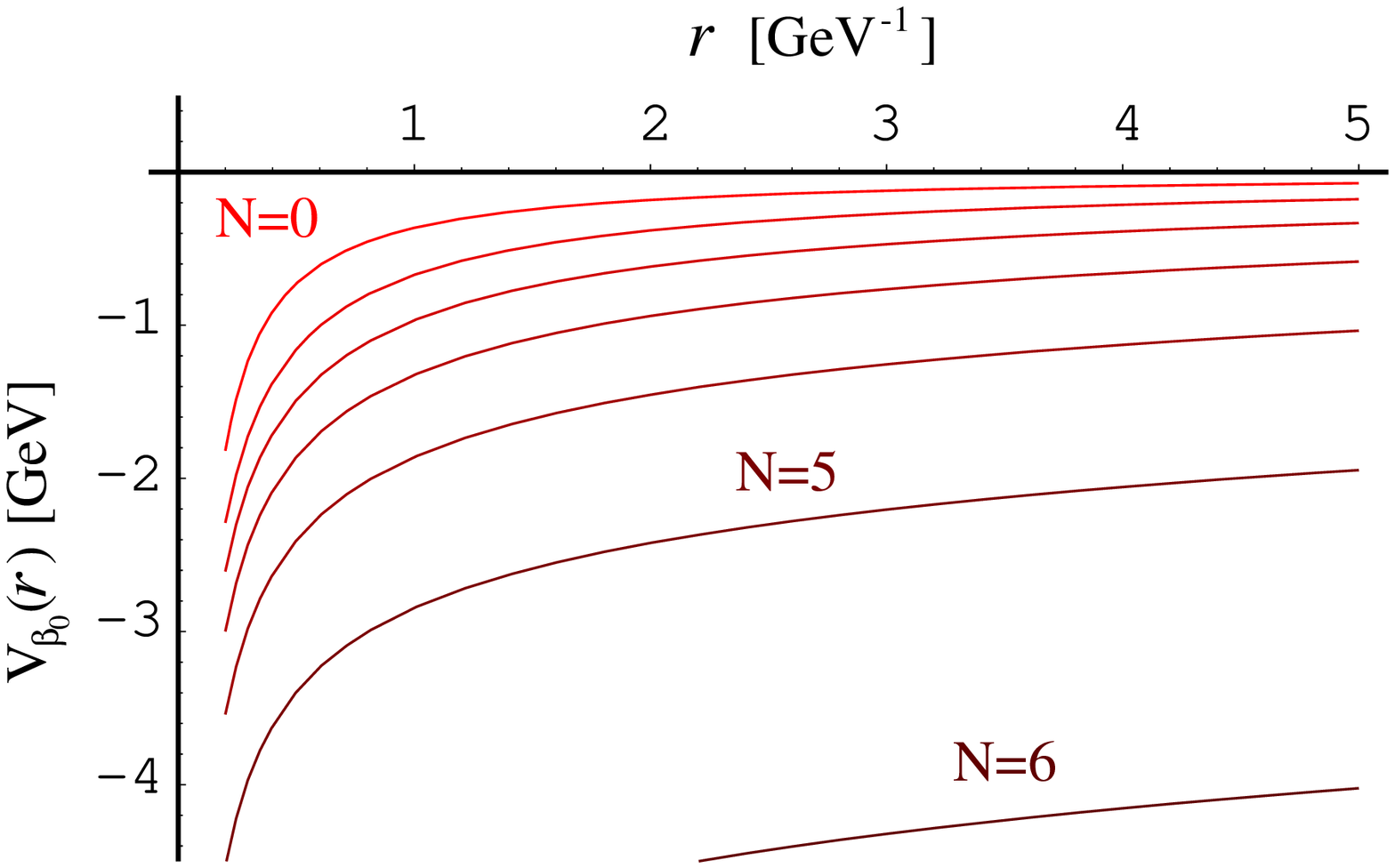}\vspace{5mm}
\hspace*{-1cm}(a)
  \end{minipage}
  \hspace*{\fill}
  \begin{minipage}{5.0cm}\centering
    \hspace*{-1cm}
    \includegraphics[width=8cm]{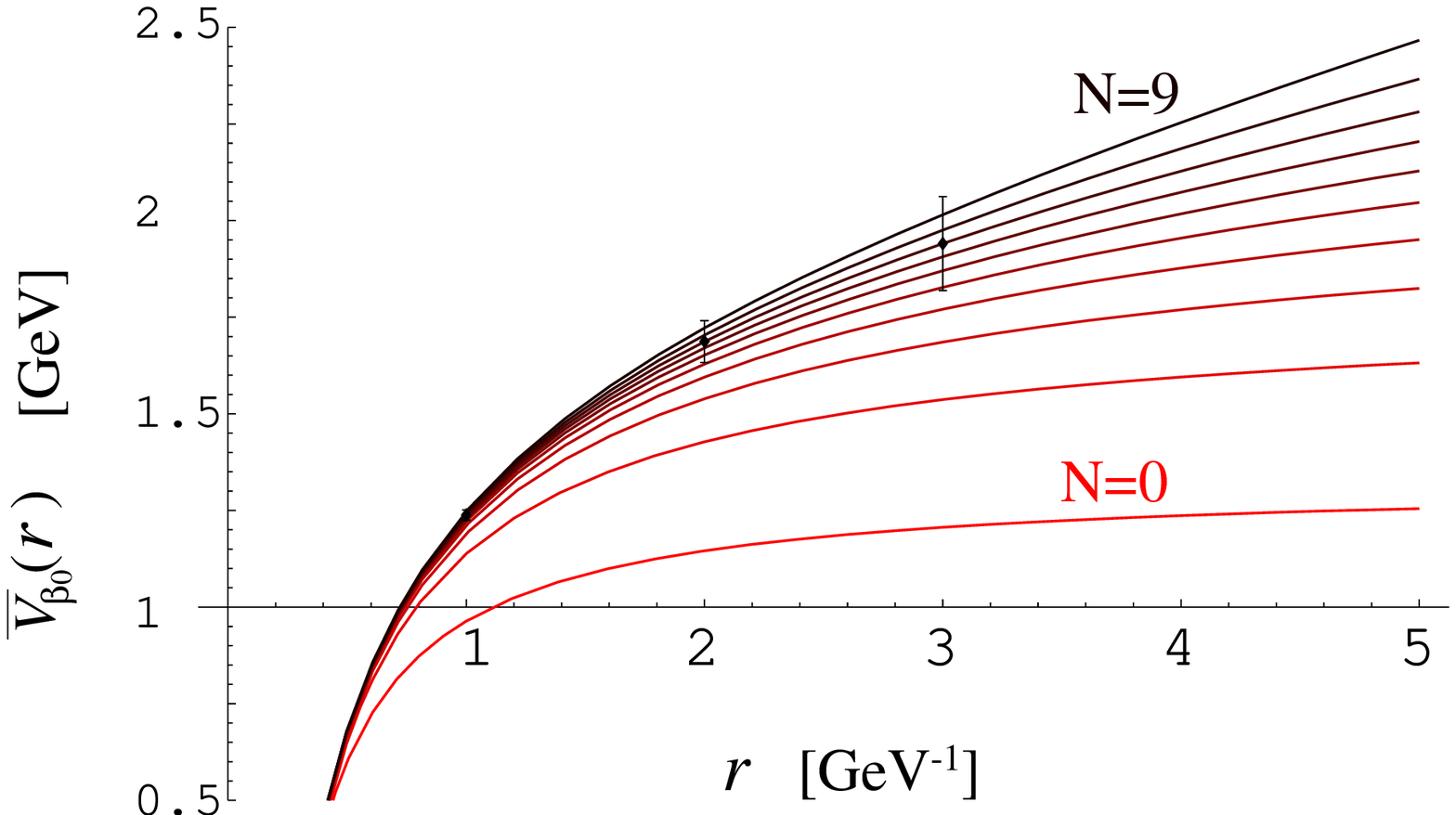}\vspace{9mm}
    \hspace*{1.2cm}(b)
  \end{minipage}
  \hspace*{\fill}
  \\
  \hspace*{\fill}
\caption{\footnotesize
The QCD potential in the large-$\beta_0$ approximation
truncated at $O(\alpha_S^{N+1})$ term.
We set $\mu=2.49$~GeV, $n_l=4$ and $\alpha_S(\mu)=0.273$
[corresponding to $\alpha_S^{(5)}(M_Z)=0.1181$].
(a) Before subtraction of the leading renormalon.
(b) After subtraction of the leading renormalon.
      \label{large-beta0}
}
  \hspace*{\fill}
\end{figure}
In Fig.~\ref{large-beta0}(a) we show
the QCD potential in the large-$\beta_0$ approximation truncated
at the $(N \! + \! 1)$-th term, $\sum_{n=0}^N V^{(n)}_{\beta_0}(r)$,
for $N=0,1,2,\dots$ and $n_l=4$.
We see that the higher order corrections are indeed large and
almost constant (independent of $r$).

It was found \cite{renormalon1,renormalon2} 
that the leading renormalon contained in the QCD potential
gets cancelled in the total energy of a static quark-antiquark pair,
\bea
E_{\rm tot}(r) \equiv 2 m_{\rm pole} + V_{\rm QCD}(r) ,
\label{totene}
\eea
if the pole mass $m_{\rm pole}$ is expressed in terms of
the $\overline{\rm MS}$ mass.
Namely, when expressed in terms of the $\overline{\rm MS}$ mass and
in a series expansion in $\alpha_S(\mu)$,
the pole mass contains the leading renormalon \cite{bbb}
which is one half in size
and opposite in sign of the leading renormalon of $V_{\rm QCD}(r)$.
Thus, the total energy $E_{\rm tot}(r)$ is
free from the leading renormalon uncertainties.
$E_{\rm tot}(r)$ possesses a residual uncertainty originating from the
next-to-leading renormalon \cite{al},
\bea
\delta E_{\rm tot}(r) \sim \Lambda \times ( \Lambda r )^2 ,
\eea
which is smaller than the leading renormalon uncertainty
in the range $r \simlt \Lambda^{-1}$.
Shown in Fig.~\ref{large-beta0}(b) is
the QCD potential in the large-$\beta_0$ approximation
[truncated at the $(N \! + \! 1)$-th term] after 
the leading renormalon is subtracted at each order of $\alpha_S(\mu)$:
\bea
\overline{V}_{\beta_0}(r) 
= \sum_{n=0}^\infty 
\left[ V^{(n)}_{\beta_0}(r)
+ C_F \, 4 \pi \alpha_S(\mu) \cdot \frac{\mu \, e^{5/6}}{2\pi^2} \cdot
\biggl\{ \frac{\beta_0 \alpha_S(\mu)}{2\pi} \biggr\}^n \,
\, n! \,
\right] .
\label{subt}
\eea
One sees that the series expansion of the potential has become 
much more convergent as compared to Fig.~\ref{large-beta0}(a).
For a particular choice of the scale $\mu = 2.49$~GeV, the term 
on the right-hand-side of Eq.~(\ref{subt})
becomes
smallest at around $n = 7$ in the range 
$1~{\rm GeV}^{-1} < r < 5~{\rm GeV}^{-1}$.
Hence, the error bars corresponding to the 
next-to-leading renormalon uncertainty
$\pm {\frac{1}{2}}
\Lambda \cdot ( \Lambda r )^2$ (taking $\Lambda = 300$~MeV) are attached
to the potential for $N=7$ in the same figure.
We may consider that the line for $N=7$ together with the error bars 
indicate a typical accuracy 
of the perturbative prediction for the QCD potential,
when the leading renormalon is cancelled.
We see that the potential is bent upwards at long distances as compared
to the leading Coulomb potential ($N=0$).
If we choose a smaller scale for $\mu$, the term becomes smallest at a
smaller $n$.
In this case, convergence properties become better at larger $r$,
where we obtain a value of $\overline{V}_{\beta_0}(r)$
consistent with $N=7$ of Fig.~\ref{large-beta0}(b)
with less terms (smaller $N$).
Similarly to the leading renormalon case, 
the uncertainty is $\mu$-independent, nonetheless.

\section{The Total Energy of a $q\bar{q}$ System}

Now we examine the total energy of a quark-antiquark pair,
defined in Eq.~(\ref{totene}), exactly up to 
${\cal O}(\alpha_S^3)$.
This quantity is free from the leading renormalon uncertainty;
in fact the cancellation of the leading renormalons occurs at a
deeper level than what can be seen in the large-$\beta_0$ 
approximation \cite{renormalon2}.
We also note that the cancellation at each order of 
perturbative expansion
is realized only when we use the same coupling constant in expanding
$m_{\rm pole}$ and $V_{\rm QCD}(r)$.\footnote{
This can be seen, for example, from the fact that the order 
$n_0 = 2\pi/(\beta_0 \alpha_S(\mu))$
at which Eq.~(\ref{asympt}) becomes smallest
is dependent on the value of $\alpha_S(\mu)$ used for the expansion.
}

The QCD potential of the theory with $n_l$ massless flavors only\footnote{
The QCD potential of the theory which contains
$n_h$ heavy flavors (with mass $m$) and $n_l$ massless flavors
coincides with the potential in Eq.~(\ref{QCDpot}) up to
${\cal O}(\alpha_S^3)$ if we count $1/r = {\cal O} (\alpha_S m)$
and if we properly match the coupling to that of the theory with
$n_l$ massless flavors only.
}
is given, up to ${\cal O}(\alpha_S^3)$, by 
\bea
V_{\rm QCD}(r) & = &
- \, C_F \frac{\alpha_S(\mu)}{r}
\Biggl[ \, 1 + 
\biggl( \frac{\alpha_S(\mu)}{4\pi} \biggr) 
  ( 2 \beta_0 \ell + a_1 )
\nonumber \\ && ~~~~~~~~~~~~~~
+
\biggl( \frac{\alpha_S(\mu)}{4\pi} \biggr)^2
  \left\{ \beta_0^2 \Bigl( 4 \ell^2  + \frac{\pi^2}{3} \Bigr)
  + 2 ( \beta_1 + 2 \beta_0 a_1 ) \ell + a_2 \right\}
\Biggr] ,
\label{QCDpot}
\eea
where \cite{ps}
\bea
&&
\ell = \log (\mu r) + \gamma_E ,
\\ &&
\beta_0 = 11 - \frac{2}{3} n_l, 
~~~~~~
\beta_1 = 102 - \frac{38}{3} n_l,
\\ &&
a_1 = \frac{31}{3} - \frac{10}{9} n_l ,
~~~~~
a_2 = {\frac{4343}{18}} + 36\,{{\pi }^2} +   66\,{\zeta_3} - 
  {\frac{9\,{{\pi }^4}}{4}} - {n_l}\,
   \left( {\frac{1229}{27}} + {\frac{52\,{\zeta_3}}{3}} \right)  
+ {\frac{100}{81}} \,{n_l^2} .
\nonumber \\
\eea
The relation between the pole mass and the $\overline{\rm MS}$ mass
has been computed up to three loops in a full theory,
which contains $n_h$ heavy flavors and $n_l$ massless flavors \cite{mr}.
(The same relation was obtained numerically in \cite{chst} in a certain 
approximation.)
Rewriting the relation in terms of the coupling of the
theory with $n_l$ massless flavors only, we find\footnote{
When $n_h=1$,
this relation coincides with Eq.(14) of \cite{mr}, which
is given numerically (indirectly through $\beta_0$ 
and $\beta_1$).
Note that,  in the other
formulas of \cite{mr}, the coupling of the full theory is used.
}
\bea
m_{{\rm pole}} = \overline{m} \left\{ 1 + {4\over 3}\,  
{\alpha_S(\overline{m})\over \pi} 
+   \left({\alpha_S(\overline{m})\over \pi}\right)^2 d_1 
+  \left({\alpha_S(\overline{m})\over \pi}\right)^3 d_2 \right\},
\eea
where
$\overline{m} \equiv m_{\overline{\rm MS}}(m_{\overline{\rm MS}})$
denotes the renormalization-group-invariant $\overline{\rm MS}$ mass,
and
\bea
d_1 &=& {\frac{3049}{288}} + {\frac{2\,{{\pi }^2}}{9}} 
+ {\frac{{{\pi }^2}\,\log 2}{9}} - 
  {\frac{{\zeta_3}}{6}} + 
  {n_l}\,\left( -{\frac{71}{144}} - {\frac{{{\pi }^2}}{18}} \right)
      + {n_h}\,\left( -{\frac{143}{144}} + 
     {\frac{{{\pi }^2}}{9}} \right)  ,
\\
d_2 &= &
{\frac{1145453}{10368}}  + 
    {\frac{25379\,{{\pi }^2}}{2592}} + 
    {\frac{235\,{{\pi }^2}\,\log 2}{54}} - {\frac{9\,\zeta_3}{8}} 
- {\frac{341\,{{\pi }^4}}{2592}} - 
    {\frac{7\,{{\pi }^2}\,{\log^2 2}}{27}} 
\nonumber \\&&
-   {\frac{19\,{\log^4 2}}{54}} - {\frac{76\,a_4}{9}}
-   {\frac{1331\,{{\pi }^2}\,\zeta_3}{432}} 
+   {\frac{1705\,\zeta_5}{216}}
\nonumber \\&&
+ n_l \Biggl(
-{\frac{81227}{7776}}  - 
    {\frac{965\,{{\pi }^2}}{648}} 
-   {\frac{11\,{{\pi }^2}\,\log 2}{81}}
- {\frac{707\,\zeta_3}{216}
+ {\frac{61\,{{\pi }^4}}{1944}}  
+   {\frac{2\,{{\pi }^2}\,{\log^2 2}}{81}} + 
    {\frac{{\log^4 2}}{81}} + {\frac{8\,a_4}{27}} }
\Biggr)
\nonumber \\&&
+n_h \Biggl(
   -{\frac{157007}{7776}} + 
    {\frac{13627\,{{\pi }^2}}{1944}} - 
    {\frac{640\,{{\pi }^2}\,\log 2}{81}} 
+   {\frac{751\,\zeta_3}{216}}
+ {\frac{41\,{{\pi }^4}}{972}} - 
    {\frac{{{\pi }^2}\,{\log^2 2}}{81}} + {\frac{{\log^4 2}}{81}}   
\nonumber \\&&
~~~~~ ~~~
+ {\frac{8\,a_4}{27}} - 
    {\frac{{{\pi }^2}\,\zeta_3}{4}} + 
    {\frac{5\,\zeta_5}{4}}
\Biggr)
\nonumber \\&&
+n_l^2 \Biggl(
{\frac{2353}{23328}} + {\frac{13\,{{\pi }^2}}{324}} + 
    {\frac{7\,\zeta_3}{54}}
\Biggr)
+ n_l \, n_h \Biggl(
   {\frac{5917}{11664}} - {\frac{13\,{{\pi }^2}}{324}} - 
    {\frac{2\,\zeta_3}{27}}
\Biggr)
\nonumber \\&&
+n_h^2 \Biggl(
   {\frac{9481}{23328}} - {\frac{4\,{{\pi }^2}}{405}} - 
    {\frac{11\,\zeta_3}{54}}
\Biggr) ,
\eea
with $a_4 = {\rm Li}_4(\frac{1}{2})$.
Furthermore, we rewrite $\alpha_S(\overline{m})$ in terms of
$\alpha_S(\mu)$ using the renormalization-group evolution of the
coupling constant.
Thus, we examine the series expansion of
$E_{\rm tot}(r;\overline{m},\alpha_S(\mu))$ in $\alpha_S(\mu)$ up to 
${\cal O}(\alpha_S^3)$.
Qualitatively the series shows a convergence property very similar to
$\overline{V}_{\beta_0}(r)$ for $N=0,1,2$; cf.\ Fig.~\ref{large-beta0}(b).

The obtained total energy depends on the scale $\mu$ due to 
truncation of the series at a finite order.
One finds that, when $r$ is small, the series
converges better and the value of $E_{\rm tot}(r)$ is less $\mu$-dependent
if we choose a large scale for $\mu$, whereas when $r$ is larger,
the series converges better and the value of $E_{\rm tot}(r)$ is 
less $\mu$-dependent if we choose a smaller scale for $\mu$.
Taking into account this property, 
we will fix the scale $\mu$ in two different ways below:
\begin{enumerate}
\item
We fix the scale $\mu = \mu_1(r)$ by demanding stability against variation of 
the scale:
\bea
\left. \mu \frac{d}{d\mu} E_{\rm tot}(r;\overline{m},\alpha_S(\mu))
\right|_{\mu = \mu_1(r)} = 0 .
\label{scalefix1}
\eea
\item
We fix the scale $\mu = \mu_2(r)$
on the minimum of the absolute value of the last known term 
[${\cal O}(\alpha_S^3)$ term] of $E_{\rm tot}(r)$:
\bea
\left. \mu \frac{d}{d\mu}
\Bigl[ E^{(3)}_{\rm tot}(r;\overline{m},\alpha_S(\mu)) \Bigr]^2 \,
\right|_{\mu = \mu_2(r)} = 0 .
\label{scalefix2}
\eea
\end{enumerate}

In this analysis we examine the total energy of a $b\bar{b}$ system.
We set 
$\overline{m}_b \equiv 
m_b^{\overline{\rm MS}}(m_b^{\overline{\rm MS}})
= 4.203$~GeV, which is taken from \cite{bsv1}.
(Its error is estimated to be about $\pm 30$~MeV.)
For simplicity we analyze $E_{\rm tot}(r)$ in 
two hypothetical cases:
(i) when $m_c = 0$ ($n_l=4$ and $n_h=1$), and
(ii) in the limit $m_c \to m_b$ ($n_l=3$ and $n_h=2$).
The real world lies somewhere in between the two cases:
the charm quark decouples in the excited states of bottomonium
but not in the ground state \cite{bsv2}.
A more precise analysis requires inclusion of nonzero $m_c$ effects
into $E_{\rm tot}(r)$,
which will be reported elsewhere \cite{elsewhere}.
The input value of the strong coupling constant
is $\alpha_S^{(5)}(M_Z) = 0.1181$ \cite{pdg}.
We evolve the coupling and match 
it to the couplings of the theory with $n_l=4$ and 3 successively 
by solving the renormalization-group equation numerically with
the 3-loop beta function and by
using the 3-loop matching condition \cite{lrv}\footnote{
We take the matching scales as $\overline{m}_b$ and 
$\overline{m}_c (= \overline{m}_b)$, 
respectively.} 
(3-loop running).

\begin{figure}[tbp]
  \hspace*{\fill}
  \begin{minipage}{5.0cm}\centering
    \hspace*{-2.3cm}
    \includegraphics[width=8cm]{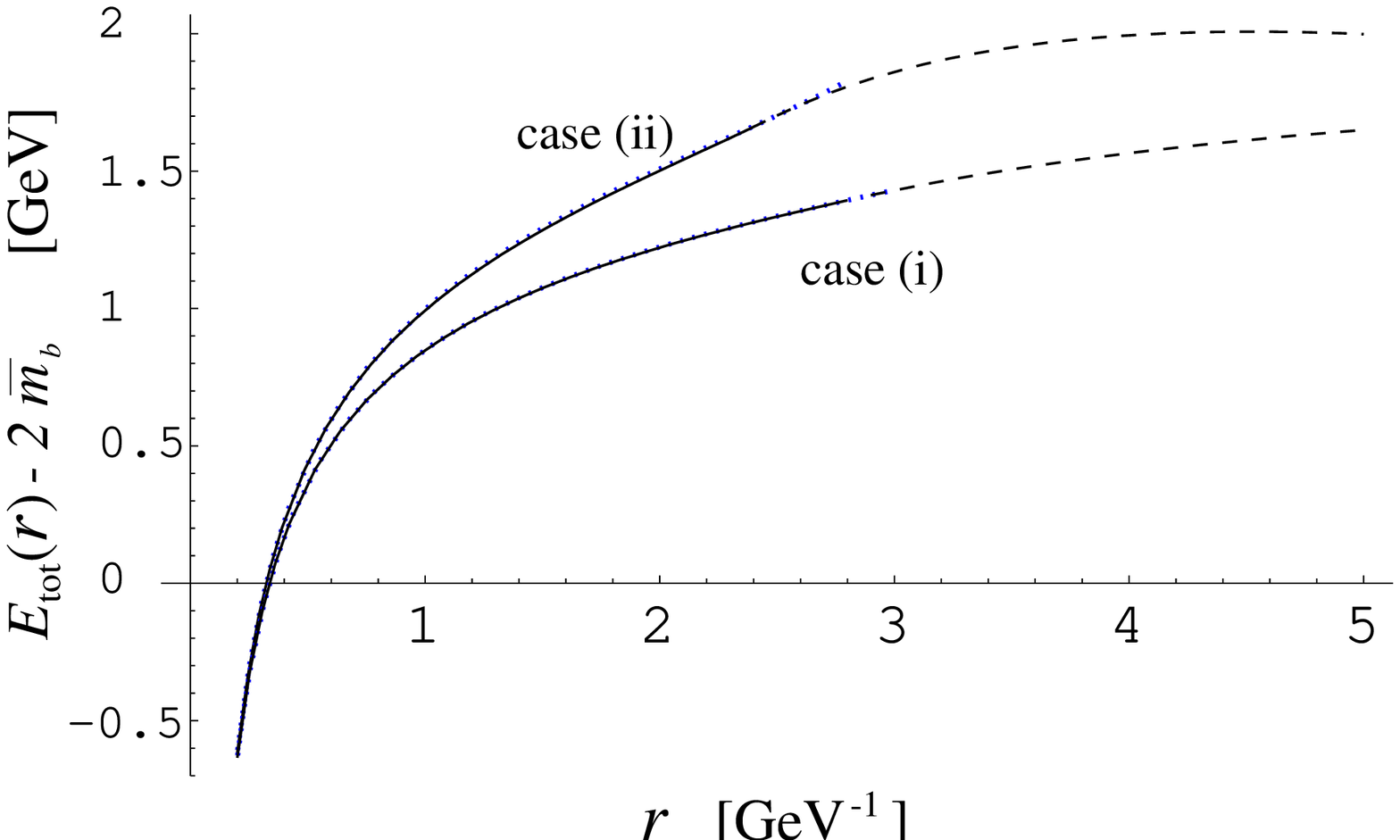}\vspace{3mm}
\hspace*{-1cm}(a)
  \end{minipage}
  \hspace*{\fill}
  \begin{minipage}{5.0cm}\centering
    \vspace*{-7mm}\hspace*{-1cm} 
    \includegraphics[width=8cm]{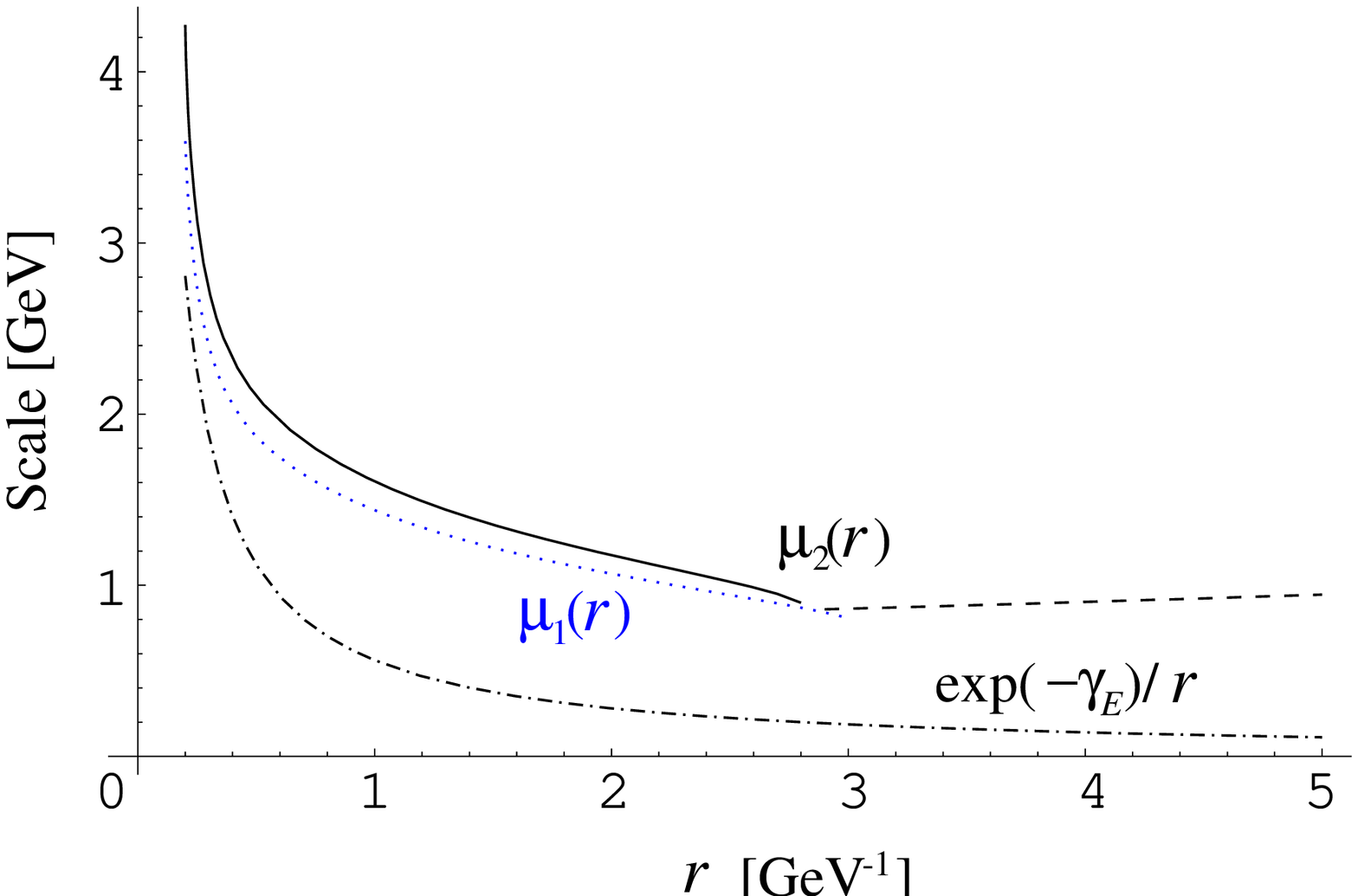}\vspace{5mm}
    \hspace*{1.2cm}(b)
  \end{minipage}
  \hspace*{\fill}
  \\
  \hspace*{\fill}
\caption{\footnotesize
(a) The total energy of a $b\bar{b}$ system 
measured from $2 \overline{m}_b$
in two hypothetical cases. 
In each case, the scale is fixed by $\mu = \mu_1(r)$ (dotted lines) or
$\mu = \mu_2(r)$ (solid lines if $E_{\rm tot}^{(3)}(r) = 0$;
dashed lines if $|E_{\rm tot}^{(3)}(r)| > 0$).
(b) The scales chosen by the scale-fixing prescriptions
(\ref{scalefix1}) and (\ref{scalefix2}) in case (i).
The notations are same as in (a).
A conventional scale choice $\mu = \exp (-\gamma_E)/r$ is
also shown (dotdashed line).
      \label{totalenergy}
}
  \hspace*{\fill}
\end{figure}
Fig.~\ref{totalenergy}(a) 
shows $E_{\rm tot}(r)$ (measured from $2 \overline{m}_b$) 
for the two cases (i) and (ii).
In each case $E_{\rm tot}(r)$ are plotted with  the
two different scale-fixing prescriptions;
the total energy hardly changes whether we choose $\mu=\mu_1(r)$
or $\mu=\mu_2(r)$.
In case (i), the minimal sensitivity scale $\mu_1(r)$ exists only in
the range $r \simlt 3$~GeV$^{-1}$;
for the choice $\mu=\mu_2(r)$,
the minimum value of $|E^{(3)}_{\rm tot}(r)|$ is zero in the range
$r \simlt 3$~GeV$^{-1}$, whereas $|E^{(3)}_{\rm tot}(r)|>0$
in the range $r \simgt 3$~GeV$^{-1}$.
These features indicate an instability of the perturbative
prediction for $E_{\rm tot}(r)$ at $r \simgt 3$~GeV$^{-1}$.
The scales $\mu_1(r)$ and $\mu_2(r)$ are shown as functions of $r$
in Fig.~\ref{totalenergy}(b).
For comparison, we also show $\mu = \exp (-\gamma_E)/r$, which 
has been considered as a natural scale of the QCD potential,
$V_{\rm QCD}(r)$, conventionally.
One sees that $\mu_1(r)$ and $\mu_2(r)$ are considerably larger than
$\exp (-\gamma_E)/r$.
The scales chosen in case (ii) are similar.
\begin{table}
\begin{center}
\begin{tabular}{l||r|r|r||r|r|r}
\hline
& \multicolumn{6}{c}{case (i)}\\
\cline{2-7}
& \multicolumn{3}{c||}{$\mu = \mu_1(r)$} 
& \multicolumn{3}{c}{$\mu = \mu_2(r)$} \\
\cline{2-7}
& $E_{\rm tot}^{(1)}(r)$ & $E_{\rm tot}^{(2)}(r)$ & $E_{\rm tot}^{(3)}(r)$ 
& $E_{\rm tot}^{(1)}(r)$ & $E_{\rm tot}^{(2)}(r)$ & $E_{\rm tot}^{(3)}(r)$ \\
\hline
$r=1~{\rm GeV}^{-1}$ & 797 & 69 & $-17$ & 750 & 98 & 0 \\
\hline
$r=2~{\rm GeV}^{-1}$ & 1255 & $-14$ & $-17$ & 1173 & 48 & 0 \\
\hline
$r=3~{\rm GeV}^{-1}$ & 1709 & $-290$ & $13$ & 1606 & $-185$ & 9 \\
\hline
\hline
& \multicolumn{6}{c}{case (ii)}\\
\cline{2-7}
& \multicolumn{3}{c||}{$\mu = \mu_1(r)$} 
& \multicolumn{3}{c}{$\mu = \mu_2(r)$} \\
\cline{2-7}
& $E_{\rm tot}^{(1)}(r)$ & $E_{\rm tot}^{(2)}(r)$ & $E_{\rm tot}^{(3)}(r)$ 
& $E_{\rm tot}^{(1)}(r)$ & $E_{\rm tot}^{(2)}(r)$ & $E_{\rm tot}^{(3)}(r)$ \\
\hline
$r=1~{\rm GeV}^{-1}$ & 962 & 70 & $-32$ & 879 & 117 & 0 \\
\hline
$r=2~{\rm GeV}^{-1}$ & 1659 & $-116$ & $-31$ & 1502 & 4 & 0 \\
\hline
$r=3~{\rm GeV}^{-1}$ & -- & -- & -- & 1994 & $-197$ & 70 \\
\hline
\end{tabular}
\end{center}
\vspace{-3mm}
\caption{\footnotesize
Series expansion of the total energy in $\alpha_S(\mu)$ with
the two scale choices Eqs.~(\ref{scalefix1}) and (\ref{scalefix2}).
$E_{\rm tot}^{(n)}(r)$ denotes the ${\cal O}(\alpha_S^n)$ term of
$E_{\rm tot}(r)$.
All numbers are in MeV unit.
The minimal sensitivity scale $\mu_1(r)$ exists only at 
$r < 2.8~{\rm GeV}^{-1}$ in case (ii).
}
\label{tab}
\end{table}
In Table~\ref{tab} we show each term of the series expansion of
$E_{\rm tot}(r)$.
The series shows healthy convergent behavior at 
$r \simlt 3~{\rm GeV}^{-1}$.

At this stage, let us discuss why the scales 
$\mu_1(r)$ and $\mu_2(r)$ are considerably larger than
$\exp (-\gamma_E)/r$.
For this purpose we use an approximate expression for the
pole mass, which
follows from the fact that the dominant contribution to
the pole-$\overline{\rm MS}$ mass relation can be read from the infrared 
region, 
loop momenta $q \ll \overline{m}$, of the QCD static 
potential \cite{renormalon2}: 
\bea
&&
2m_{\rm pole} \approx 2 \overline{m} +
{\hbox to 18pt{ \hbox to -5pt{$\displaystyle \int$} 
\raise-19pt\hbox{$\scriptstyle |\vec{q}|< \overline{m}$} }}
\frac{d^3\vec{q}}{(2\pi)^3} \, | V_{\rm QCD}(q) |
= 2 \overline{m} + \frac{2 C_F}{\pi} \int_0^{\overline{m}} dq \, 
\tilde{\alpha}_V (q) .
\label{poleapprox} 
\eea 
Here, $V_{\rm QCD}(q) = - C_F 4\pi \tilde{\alpha}_V(q)/q^2$ is the QCD static 
potential in momentum space.
Then the total energy can be written approximately as
\bea
E_{\rm tot}(r) & \approx  &
2 \overline{m} + 
\int \frac{d^3\vec{q}}{(2\pi)^3} \, | V_{\rm QCD}(q) | \,
\Bigl[ \, \theta (\overline{m}-q) - \exp (i \vec{q} \cdot \vec{r}) \,
\Bigr]
\\&=&
2 \overline{m} + 
\frac{2 C_F}{\pi} \int_0^{\infty} dq \, 
\tilde{\alpha}_V (q) \,
\left[ \theta (\overline{m}-q) - \frac{\sin (qr)}{qr}
\right] .
\label{average}
\eea
In the integrands, the factors in the brackets
$[ \cdots ]$ are appreciable only in the range
$1/r \simlt q < \overline{m}$.
So, roughly speaking, $E_{\rm tot}(r)$ is determined from an average 
$\left<  \tilde{\alpha}_V \right>$ of the $V$-scheme coupling
$\tilde{\alpha}_V (q)$ over the range $1/r \simlt q < \overline{m}$.
When evaluating this quantity in fixed-order perturbation theory,
a scale $\mu(r)$ which represents this average coupling, i.e.
$\tilde{\alpha}_V(\mu(r)) \approx \left<  \tilde{\alpha}_V \right>$, would
be a most natural scale.
Such a scale should lie between $1/r$ and $\overline{m}$.
This argument is in contrast with the conventional
principle for the scale choice for the QCD potential $V_{\rm QCD}(r)$.
Apart from $\Lambda_{\rm QCD}$, the QCD potential contains only one scale
$1/r$, so that the choice of scale has been almost automatic, $\mu \sim 1/r$.
The potential alone, however, has a large uncertainty due to the leading
renormalon.
It stems from the contribution of $\tilde{\alpha}_V (q)$ at
$q \sim \Lambda_{\rm QCD}$.
On the other hand, the total energy is free from the leading renormalons
by cutting out large contributions from $\Lambda_{\rm QCD} \sim q < 1/r$
as seen in Eq.~(\ref{average}).
Consequently the relevant scale is shifted to higher momentum
region in comparison to that of $V_{\rm QCD}(r)$.

It would also be instructive to compare the above scale choices with
the Brodsky-Lepage-Mackenzie (BLM) scale-fixing prescription \cite{blm}
applied to $V_{\rm QCD}(r)$ and $E_{\rm tot}(r)$, respectively.
In this prescription (at the lowest order),
the part of higher-order corrections to $V_{\rm QCD}(r)$ or to
$E_{\rm tot}(r)$
given by the large-$\beta_0$ approximation is absorbed into the
scale choice.
For the QCD potential,
at the lowest order the BLM scale is fixed as
$\mu = \exp ( - \frac{5}{6} - \gamma_E )/r 
\approx 0.43 \,  \exp ( - \gamma_E )/r$.
For the total energy, the BLM scale at the lowest order is given by
$\mu = f(\overline{m}r)/r$, where
\bea
f(x) = \exp
\left[
\Bigl( \log x - \frac{\pi^2}{8} + \gamma_E - \frac{53}{192} \Bigr)
\frac{x}{x-\frac{\pi}{2}}
- \frac{5}{6} - \gamma_E
\right] .
\eea
Due to the singularity of $f(x)$ at $x = \pi/2$,
the BLM scale turns out to be unstable around $r = \pi/(2\overline{m})$.
This is because the coefficient of $\beta_0 \log \mu$ 
in $E_{\rm tot}(r;\overline{m},\alpha_S(\mu))$ 
becomes small by a cancellation
between $V_{\rm QCD}(r)$ and $2 m_{\rm pole}$.
In this region of $r$, the BLM prescription for $E_{\rm tot}(r)$
would be unreliable.
For $r > \pi/(2\overline{m})$, the function
$f(\overline{m}r)$ increases monotonically.
Setting $\overline{m}=4.203$~GeV, we find that 
at $r \simgt 0.6~{\rm GeV}^{-1}$ 
the scale $\mu = f(\overline{m}r)/r$ exceeds the BLM scale of the QCD
potential; at $r \simgt 1~{\rm GeV}^{-1}$, $\mu = f(\overline{m}r)/r$ 
becomes almost independent of $r$, $\mu \approx 0.5$~GeV, converging
towards the BLM scale of the pole mass.
These features in the region $r \simgt 0.6~{\rm GeV}^{-1}$
are consistent with the
results of the analysis given in Sec.~\ref{sec:largebeta0}.
Since the higher-order corrections are large for $V_{\beta_0}(r)$,
the BLM scale of $V_{\rm QCD}(r)$ tends to be small.
On the other hand, since the higher-order corrections are smaller for
$\overline{V}_{\beta_0}(r)$,
the BLM scale of $E_{\rm tot}(r)$ is larger.
Intuitively the BLM scale of a quantity sensitive to renormalons
is attracted towards the $\Lambda_{\rm QCD}$ scale, whereas
that of a renormalon-free quantity is determined by a short-distance
scale.
Thus, the qualitative features of the BLM scales
agree with those of the scales shown
in Fig.~\ref{totalenergy}(b) in the range 
$1~{\rm GeV}^{-1} \simlt r \simlt 5~{\rm GeV}^{-1}$, although
the level of agreement is not very accurate.
At shorter distances $r \approx \pi/(2\overline{m})$, 
validity of the BLM prescription for the total energy
seems doubtful on account of the large cancellation.

We return to the discussion of $E_{\rm tot}(r)$.
\begin{figure}[tbp]
  \hspace*{\fill}
    \includegraphics[width=10cm]{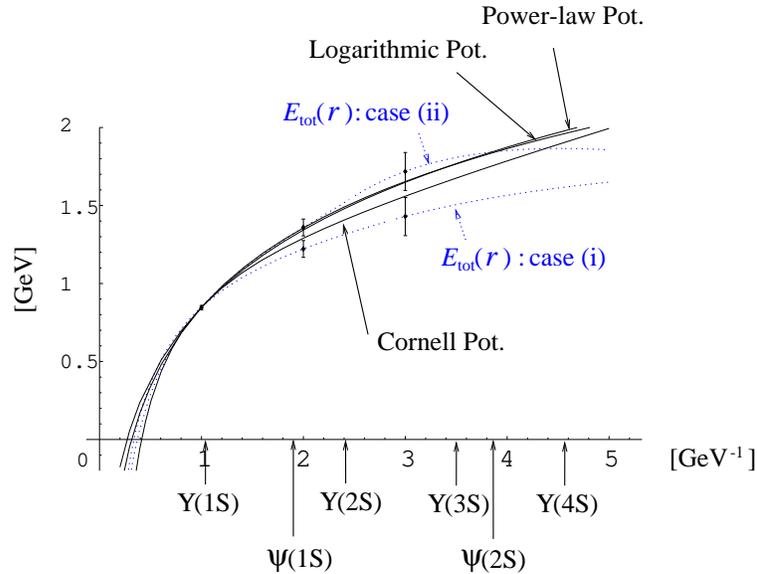}
  \hspace*{\fill}
  \\
  \hspace*{\fill}
\caption{\footnotesize
A comparison of the total energy of a $b\bar{b}$ system in the
two hypothetical cases (dotted lines) and typical phenomenological
potentials (solid lines).
For a reference, we show typical sizes of the bottomonium and charmonium
$S$ states as determined from the r.m.s.\ interquark distances with respect to
the Cornell potential: $\sqrt{\left< r^2 \right>}_{\rm Cornell}$.
      \label{compare}
}
  \hspace*{\fill}
\end{figure}
In Fig.~\ref{compare} we compare the total energies in case (i) and (ii) with
typical phenomenological potentials used in phenomenological
approaches.
We take:
\begin{itemize}
\item
A Coulomb-plus-linear potential (Cornell potential) \cite{cornell}:
\bea
V(r) = - \frac{\kappa}{r} + \frac{r}{a^2}
\eea
with $\kappa = 0.52$ and $a = 2.34$~GeV$^{-1}$.
\item
A power-law potential \cite{martin}:
\bea
V(r) = - 8.064~{\rm GeV} +
(6.898~{\rm GeV})(r\times 1~{\rm GeV})^{0.1} .
\eea
\item
A logarithmic potential \cite{qr}:
\bea
V(r) = -0.6635~{\rm GeV} + (0.733~{\rm GeV}) \log (r\times 1~{\rm GeV}) .
\eea
\end{itemize}
We may consider the differences of these potentials in the
range $0.5~{\rm GeV}^{-1} \simlt r \simlt 5~{\rm GeV}^{-1}$
as uncertainties of the phenomenologically determined potentials.
In order to make a clear comparison, arbitrary constants have been added
to all the potentials and $E_{\rm tot}(r)$
such that their values coincide at $r=1$~GeV$^{-1}$.
As stated, we expect the perturbative prediction for a realistic
$E_{\rm tot}(r)$ to lie between those for the cases (i) and (ii).
It appears to be in good agreement with the phenomenological
potentials in the above range.
The level of agreement is consistent with
the uncertainties expected from
the next-to-leading renormalon contributions (indicated by the
error bars).

\section{The Interquark Force}
\setcounter{footnote}{0}

Instead of the total energy, we may also consider the interquark force
defined by
\bea
F(r) &\equiv& 
- \frac{d}{dr} E_{\rm tot}(r) = - \frac{d}{dr} V_{\rm QCD}(r)
\\
& \equiv &
- C_F \frac{\alpha_F (1/r)}{r^2} . 
\eea
The last line defines the ``$F$-scheme'' coupling constant
$\alpha_F(\mu)$.
The interquark force is also free from the leading renormalon.
In fact it has been noted \cite{melles} that the perturbative expansion
of $F(r)$ is more convergent than that of the potential $V_{\rm QCD}(r)$.
Since in the zero quark mass limit $F(r)$ is dependent only on $r$, 
we may determine
its $r$-dependence using the renormalization-group equation:
\bea
\mu^2 \frac{d}{d\mu^2} \alpha_F (\mu) = \beta_F (\alpha_F) .
\eea
It is instructive to compare the beta functions for the couplings defined
in the three different schemes.
For $n_l=4$, we find
\bea
\begin{array}{lcclclclclcl}
\beta_V(\alpha_V) &=
&-&0.6631\,{\alpha_V^2} &-& 0.3251\,{\alpha_V^3} &-& 
  1.7527\,{\alpha_V^4} &+& {\cal O}(\alpha_V^5)
&~~~~&
\mbox{($V$-scheme)}
\\
\beta_F(\alpha_F) &=
&-&0.6631\,{\alpha_F^2} &-& 0.3251\,{\alpha_F^3} &-& 
  0.5861\,{\alpha_F^4} &+& {\cal O}(\alpha_F^5)
&~~~~&
\mbox{($F$-scheme)}
\\
\beta_{\overline{\rm MS}}(\alpha_S) &=
&-&0.6631\,{\alpha_S^2} &-& 0.3251\,{\alpha_S^3} &-& 
  0.2048\,{\alpha_S^4} &+& {\cal O}(\alpha_S^5)
&~~~~&
\mbox{($\overline{\rm MS}$-scheme)}
\end{array}
\label{betafns}
\eea
The first two coefficients of the beta functions are scheme-independent 
(when we neglect the quark masses).
The third coefficient of the $V$-scheme beta function is quite
large, reflecting poor convergence of $V_{\rm QCD}(r)$
due to the leading renormalons.
The third coefficient of the $F$-scheme beta function is
smaller by factor 3 due to cancellation of the leading renormalon.
The third coefficient of the $\overline{\rm MS}$-scheme beta function
is even smaller by factor 3.
This may be due to the fact that the $F$-scheme coupling still
contains the next-to-leading renormalon contributions.
From this comparison, we may conclude that it is better to analyze
$F(r)$ rather than $V_{\rm QCD}(r)$ as a physical quantity,
in perturbative analyses.\footnote{
This is valid up to the constant term of the potential, which is important
in relating the boundstate masses to the heavy quark masses.
An alternative way may be to study renormalization-group evolution
of $V_{\rm QCD}(r)$ after subtracting the leading renormalon from it
by hand [similar to $\overline{V}_{\beta_0}(r)$ of Eq.~(\ref{subt})].
}

The observed
bottomonium spectrum is qualitatively very different 
from the Coulomb spectrum.
The largest difference is that, the level spacings between consecutive 
bottomonium $nS$ states are almost constant,
whereas in the Coulomb spectrum the level spacings decrease as $1/n^2$.
When we consider effects of the QCD radiative corrections on the 
lowest-order Coulomb potential,
one may interpret that in the QCD potential,
$-C_F \alpha_V(1/r)/r$, the $V$-scheme coupling increases at long
distances, so that the potential will be bent downwards.
This is obviously a bad interpretation, because in such a case,
the level spacings among the excited states become even smaller
than those of the Coulomb spectrum.
We should rather consider the interquark force.
A better interpretation is that in 
$F(r)=-C_F \alpha_F(1/r)/r^2$, the $F$-scheme coupling increases at long
distances, and $|F(r)|$ grows correspondingly.
This means that the slope of the potential becomes steeper at
long distances.
(Its effect resembles an addition of a linearly rising potential
to the Coulomb potential.)
Accordingly the level spacings
among the excited states increase.
Thus, the effects of the radiative corrections on the level spacings are
even qualitatively reversed, whether we consider
$V_{\rm QCD}(r)$ or $F(r)$ as the physically relevant quantity.\footnote{
It is a matter of interpretaion.
One may understand the radiative corrections in the context of
$V$-scheme and require for large non-perturbative corrections to
remedy the discrepancy from the phenomenologically determined potentials
or the results of non-perturbative (lattice) calculations
(see e.g.\ \cite{bali,kko}).
Alternatively one may understand the radiative corrections in 
the context of $F$-scheme and call for much smaller non-perturbative
corrections.
}

\begin{figure}[tbp]
  \hspace*{\fill}
    \includegraphics[width=10cm]{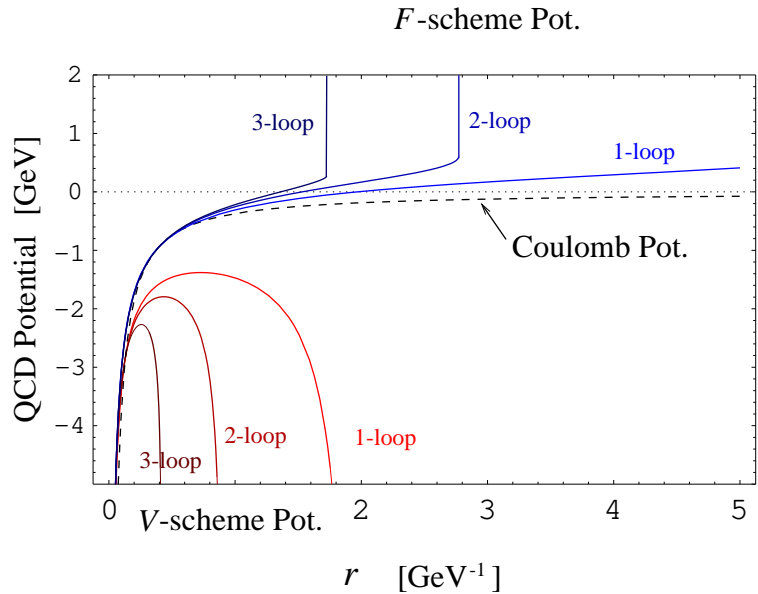}
  \hspace*{\fill}
  \\
  \hspace*{\fill}
\caption{\footnotesize
A comparison of the QCD potentials calculated in $V$-scheme and in
$F$-scheme as well as the Coulomb potential.
The Coulomb potential is given by $-C_F \alpha /r$ with
$\alpha = 0.279$.
The $V$-scheme and $F$-scheme potentials correspond to 
$\alpha_S^{(5)}(M_Z)=0.1181$.
      \label{fig-Fscheme}
}
  \hspace*{\fill}
\end{figure}
One may verify these features in Fig.~\ref{fig-Fscheme}, 
in which the Coulomb potential,
the $V$-scheme potentials and the $F$-scheme potentials are displayed.
The $V$-scheme potentials are calculated by solving the 
renormalization-group equation for $\alpha_V$ numerically,
using $\beta_V$ in Eq.~(\ref{betafns}) up to order
$\alpha_V^2$ (1-loop), order $\alpha_V^3$ (2-loop) and
order $\alpha_V^4$ (3-loop).
The $F$-scheme potentials are calculated by first solving the 
renormalization-group equation for $\alpha_F$ numerically via
$\beta_F$ in Eq.~(\ref{betafns}) and then by
integrating $-F(r)$ over $r$ numerically;
arbitrary constants are added such that the $F$-scheme potentials 
coincide the Coulomb potential at $r=0.4$~GeV$^{-1}$.
The initial values for $\alpha_V$ and $\alpha_F$ are given at
$r = \exp ( - \gamma_E ) / \overline{m}_b$ by matching to
the fixed-order results.
As can be seen, the $V$-scheme potentials become singular at fairly
short distances, $r \sim 2$~GeV$^{-1}$ (1-loop),
0.9~GeV$^{-1}$ (2-loop), and 0.4~GeV$^{-1}$ (3-loop), 
respectively.
As expected, the $F$-scheme potentials have wider ranges of 
validity\footnote{
The large discrepancy between the potentials obtained from
$\beta_V(\alpha_V)$ and $\beta_F(\alpha_F)$ was noted first in
\cite{grunberg}.
}:
they become singular at $r \sim 6.9$~GeV$^{-1}$ 
(1-loop), 2.8~GeV$^{-1}$ (2-loop),
and 1.7~GeV$^{-1}$ (3-loop), respectively.
The situation is puzzling, however, in that the predictable range reduces
as we include more terms of $\beta_F(\alpha_F)$.
The 2-loop and 3-loop $F$-scheme potentials are consistent with
the phenomenological potentials within the uncertainty
expected from the next-to-leading renormalon contributions,
in the range $0.5~{\rm GeV}^{-1} \simlt r \simlt 2.8~{\rm GeV}^{-1}$
and $0.5~{\rm GeV}^{-1} \simlt r \simlt 1.7~{\rm GeV}^{-1}$,
respectively.
On the other hand, the 1-loop $F$-scheme potential does not
satisfy this criterion.

If we take a larger input value for $\alpha_S^{(5)}(M_Z)$, the slopes of
the $F$-scheme potentials get steeper,
since $\alpha_F$ increases.
Also, it explains why $E_{\rm tot}(r)$ for case (ii) is steeper
than that for case (i) in Fig.~\ref{totalenergy}(a):
$\alpha_F$ for $n_l=3$ is larger than that for $n_l=4$ at
$r \simgt 1/\overline{m}_b$.

\section{Conclusions}

When we incorporate the cancellation of the leading renormalon
contributions,
the perturbative expansion of the total energy $E_{\rm tot}(r)$
of a $b\bar{b}$ system, up to ${\cal O}(\alpha_S^3)$ and supplemented
by the scale-fixing prescription (\ref{scalefix1}) or (\ref{scalefix2}), 
converges well at $r \simlt 3~{\rm GeV}^{-1}$.
Moreover, it agrees with the phenomenologically determined potentials
in the range 
$0.5~{\rm GeV}^{-1} \simlt r \simlt 3~{\rm GeV}^{-1}$
within the uncertainty expected from the next-to-leading renormalon
contributions.
Even at $r \simgt 3~{\rm GeV}^{-1}$, the scale-fixing prescription
(\ref{scalefix2}) gives a reasonable prediction for $E_{\rm tot}(r)$;
it appears that 
the perturbative prediction does not break down suddenly but rather the
uncertainty grows gradually as $r$ increases.
The agreement is unlikely to be accidental, since 
as soon as we take
the input $\alpha_S^{(5)}(M_Z)$ outside of the present world average
values $0.1181 \pm 0.0020$ \cite{pdg}, the agreement is lost quickly.

A non-relativistic Hamiltonian
\bea
H = 2 m_{\rm pole} + \vec{p}\, ^2/m_{\rm pole} + V_{\rm QCD}(r)
= \vec{p}\, ^2/m_{\rm pole} + E_{\rm tot}(r)
\label{simpleH}
\eea
constitutes a part of the full Hamiltonian 
(up to the order $1/c^2$)
analyzed in \cite{bsv1}. 
It detetermines the
bulk of the quarkonium level structure computed therein.
At the same time,
the above Hamiltonian is exactly the ones analyzed in the conventional
phenomenological potential-model approaches 
(at the leading order) if $E_{\rm tot}(r)$
is identified with the phenomenological potentials.
Also, it is the leading-order Hamiltonian of the more systematic
frameworks discussed in Sec.~\ref{intro}.
Thus, we find that the agreement of $E_{\rm tot}(r)$ and the
phenomenologically determined potentials
is the reason why the gross
structure of the bottomonium spectrum is reproduced well by 
the computation based on perturbative QCD.
Our observation confirms the conclusion 
of \cite{bsv1}, that once the leading renormalon contributions are cancelled,
there remain no large
non-perturbative effects, which essentially deteriorate perturbative
treatment of some of the bottomonium and charmoninum states,
but only moderate contributions comparable in size with 
the next-to-leading renormalons.

Similarly, if we analyze the interquark force $F(r)$ instead of
$V_{\rm QCD}(r)$, the range of perturbative predictability becomes 
significantly wider,
as known from the previous study \cite{melles}.
We confirm this observation using a renormalization-group analysis.
We find that
the 2-loop and 3-loop renormalization-group-improved
potentials, obtained by integrating $-F(r)$,
are consistent with the phenomenological potentials up to
$r \sim 2.8~{\rm GeV}^{-1}$ and $r \sim 1.7~{\rm GeV}^{-1}$,
respectively.

We expect that the connection elucidated in this work will be useful for 
developing deeper theoretical understandings
of the bottomonium and charmonium systems.
For more detailed comparisons, in general it would be more secure to
compute the quarkonium spectra directly rather than $E_{\rm tot}(r)$
or $F(r)$.
Indeed, the series expansions of the quarkonium energy levels
turn out to be more convergent when we include the full corrections
($\vec{p} \, ^4$-term, Darwin potential, spin-dependent potentials, etc.)\
to the ${\cal O}(1/c^2)$ Hamiltonian, as compared to the expansions
of the energy levels of the simplified Hamiltonian (\ref{simpleH})
(even after the leading renormalons are cancelled).

\section*{Acknowledgements}
The author is grateful to N.~Brambilla and A.~Vairo
for very fruitful discussions.
He also thanks S.~Recksiegel for a useful comment.

\def\app#1#2#3{{\it Acta~Phys.~Polonica~}{\bf B #1} (#2) #3}
\def\apa#1#2#3{{\it Acta Physica Austriaca~}{\bf#1} (#2) #3}
\def\npb#1#2#3{{\it Nucl.~Phys.~}{\bf B #1} (#2) #3}
\def\plb#1#2#3{{\it Phys.~Lett.~}{\bf B #1} (#2) #3}
\def\prd#1#2#3{{\it Phys.~Rev.~}{\bf D #1} (#2) #3}
\def\pR#1#2#3{{\it Phys.~Rev.~}{\bf #1} (#2) #3}
\def\prl#1#2#3{{\it Phys.~Rev.~Lett.~}{\bf #1} (#2) #3}
\def\sovnp#1#2#3{{\it Sov.~J.~Nucl.~Phys.~}{\bf #1} (#2) #3}
\def\yadfiz#1#2#3{{\it Yad.~Fiz.~}{\bf #1} (#2) #3}
\def\jetp#1#2#3{{\it JETP~Lett.~}{\bf #1} (#2) #3}
\def\zpc#1#2#3{{\it Z.~Phys.~}{\bf C #1} (#2) #3}


\begin{thebibliography}{99}

\bibitem{eq}
 E. Eichten and C. Quigg, {Phys. Rev.} {\bf D49}, 5845 (1994).

\bibitem{al}
  U. Aglietti and Z. Ligeti, Phys. Lett. {\bf B364}, 75 (1995).

\bibitem{bw}
  L.~Brown and W.~Weisberger, 
  {Phys.~Rev.} {\bf D20}, 3239 (1979).

\bibitem{ef}
  E.~Eichten and F.~Feinberg,
  {Phys.~Rev.~Lett.} {\bf 43}, 1205 (1979);
  {Phys.~Rev.} {\bf D23}, 2724 (1981).

\bibitem{gromes}
  D.~Gromes, Z.~Phys.~{\bf C22}, 265 (1984).

\bibitem{bmp}
  A.~Barchielli, E.~Montaldi and G.~Prosperi,
  Nucl.~Phys.~{\bf B296}, 625 (1988), erratum, {\bf B303}, 752;

\bibitem{bbp}
  A.~Barchielli, N.~Brambilla and G.~Prosperi,
  Nuovo Cim.~{\bf 103A}, 59 (1990).

\bibitem{pinedasoto}
  A. Pineda and J. Soto, 
  Nucl.~Phys.~Proc.~Suppl.~{\bf 64}, 428 (1998).

\bibitem{bpsv}
  N. Brambilla, A. Pineda, J. Soto and A. Vairo, 
  Nucl.~Phys. {\bf B566}, 275 (2000).

\bibitem{bsw}
  G. Bali, K.~Schilling and A.~Wachter,
  Phys. Rev. {\bf D56}, 2566 (1997).

\bibitem{ukcollab}
  UKQCD Collaboration, C.~Allton, et al.,
  Phys. Rev. {\bf D60}, 034507 (1999).

\bibitem{bsv1}
  N. Brambilla, Y. Sumino and A. Vairo, 
  Phys. Lett. {\bf B513}, 381 (2001).

\bibitem{yn3}
  S. Titard and F. Yndurain, Phys. Rev. {\bf D49}, 6007 (1994);
  Phys. Rev. {\bf D51}, 6348 (1995).

\bibitem{py}
  A. Pineda and  F. Yndur{\'a}in, 
  Phys. Rev. {\bf D58}, 094022 (1998); {\bf D61}, 077505 (2000).

\bibitem{my}
  K. Melnikov and A. Yelkhovsky, Phys. Rev. {\bf D59}, 114009 (1999).

\bibitem{bcbv}
  N. Brambilla and  A. Vairo, Phys. Rev. {\bf D62}, 094019 (2000).

\bibitem{renormalon1}
  A. Hoang, M. Smith, T. Stelzer and S. Willenbrock, Phys. Rev. {\bf D59}, 114014 (1999).

\bibitem{renormalon2}
  M. Beneke, {Phys. Lett.} {\bf B434}, 115 (1998).  

\bibitem{adm}
  T.~Appelquist, M.~Dine and I.~Muzinich,
  Phys.~Rev.~{\bf D17}, 2074 (1978).

\bibitem{pro}
  M.~Beneke, 
  hep-ph/9911490.

\bibitem{sumino}
  Y.~Sumino, 
  hep-ph/0004087.

\bibitem{bb}
  M.~Beneke and V.~Braun, Phys.~Lett.~{\bf B348}, 513 (1995).

\bibitem{bbb}
  M.~Beneke and V.~Braun, Nucl.~Phys.~{\bf B426}, 301 (1994);
  I.~Bigi, M.~Shifman, N.~Uraltsev and A.~Vainshtein, 
  Phys.~Rev.~{\bf D50}, 2234 (1994).

\bibitem{ps}
  M.~Peter, Phys. Rev. Lett.~{\bf 78}, 602 (1997); 
  Nucl. Phys. {\bf B501} 471 (1997);
  Y.~Schr\"oder, {Phys.~Lett.}~{\bf B447}, 321 (1999).  

\bibitem{mr}
  K. Melnikov and T.~v.~Ritbergen, Phys. Lett. {\bf B482}, 99 (2000).

\bibitem{chst} K. Chetyrkin and M. Steinhauser, 
  Phys. Rev. Lett. {\bf 83}, 4001 (1999); 
  Nucl. Phys. {\bf B573}, 617 (2000).

\bibitem{bsv2}
 N. Brambilla, Y. Sumino and A. Vairo, hep-ph/0108084.

\bibitem{elsewhere}
  S.~Recksiegel and Y.~Sumino, hep-ph/0109122.

\bibitem{pdg}
  D. E. Groom et al., Eur. Phys. Jour. {\bf C15}, 1 (2000).

\bibitem{blm}
  S.~Brodsky, G.~Lepage and P.~Mackenzie,
  Phys.~Rev.~{\bf D28}, 228 (1983).

\bibitem{lrv}
  S.~Larin, T.~v.~Ritbergen and J.~Vermaseren,
  Nucl.~Phys.~{\bf B438}, 278 (1995).

\bibitem{cornell}
  E.~Eichten, K.~Gottfried, T.~Kinoshita, K.~Lane and T.~Yan,
  Phys.~Rev.~{\bf D17}, 3090 (1978); {\bf D21}, 313(E) (1980);
  {\bf D21}, 203 (1980).

\bibitem{martin}
  A.~Martin, 
  Phys.~Lett.~{\bf 93B}, 338 (1980).

\bibitem{qr}
  C.~Quigg and J.~Rosner,
  Phys.~Lett.~{\bf 71B}, 153 (1977).

\bibitem{melles}
  M.~Melles,
  Phys.~Rev.~{\bf D62}, 074019 (2000).

\bibitem{bali}
  G.~Bali,
  Phys.~Rept.~{\bf 343}, 1 (2001).

\bibitem{kko}
  V.~Kiselev, A.~Kovalsky and A.~Onishchenko,
  Phys.~Rev.~{\bf D64}, 054009 (2001). 

\bibitem{grunberg}
  G.~Grunberg,
  Phys.~Rev.~{\bf D40}, 680 (1989).

\end{thebibliography}
\end{document}